\documentclass[prl,aps,amssymb,twocolumn,showpacs,floatfix]{revtex4}
\usepackage[dvips]{graphicx}
\def\be{\begin{equation}}
\def\ee{\end{equation}}
\def\bea{\begin{eqnarray}}
\def\eea{\end{eqnarray}}

\begin{document}
 
\title{Interacting electrons in disordered wires: 
Anderson localization and low-$T$ transport}
\author{I.V.~Gornyi$^{1,*}$}
\author{A.D.~Mirlin$^{1,2,\dagger}$}
\author{D.G.~Polyakov$^{1,*}$}
\affiliation{$^{1}$Institut f\"ur Nanotechnologie,
Forschungszentrum Karlsruhe, 76021 Karlsruhe, Germany \\
$^{2}$Institut f\"ur Theorie der kondensierten Materie, Universit\"at
Karlsruhe, 76128 Karlsruhe, Germany}
 
\date{May 20, 2005}

\begin{abstract} We study transport of interacting electrons in a
low-dimensional disordered system at low temperature $T$.
In view of localization by disorder, the conductivity
$\sigma(T)$ may only be non-zero due to electron-electron scattering.
For weak interactions, the
weak-localization regime crosses over with lowering $T$ into a
dephasing-induced ``power-law hopping". As $T$ is further decreased,
the Anderson localization in Fock space crucially affects $\sigma(T)$,
inducing a transition at $T=T_c$, so that
$\sigma(T<T_c)=0$. The critical behavior of $\sigma(T)$ above $T_c$ is
$\ln\sigma(T)\propto - (T-T_c)^{-1/2}$. The mechanism of
transport in the critical regime 
is many-particle transitions between distant states in Fock space.

\end{abstract}
 
\pacs{72.20.-i, 72.15.Rn, 71.30.+h, 73.63.-b}
 
\maketitle

In a pathbreaking paper \cite{anderson58}
Anderson demonstrated  that a quantum particle may become localized
by a random potential. In particular, 
in non-interacting systems of one-dimensional (1D) or
two-dimensional (2D) geometry even weak disorder
localizes all electronic states \cite{abrahams79}, 
thus leading to the exactly zero
conductivity, $\sigma(T)=0$, whatever temperature $T$. A
non-zero $\sigma(T)$ in such systems may 
only occur due to inelastic scattering processes leading to dephasing
of electrons. Two qualitatively different sources of 
dephasing are possible: (i) scattering of electrons by external 
excitations (in practice, phonons) and (ii) electron-electron (e-e)
scattering. In either case, at sufficiently high temperatures, the
dephasing rate $\tau_\phi^{-1}$ is high, so that the localization
effects are reduced to a weak-localization (WL) correction to the
Drude conductivity. This correction behaves 
as $\ln\tau_{\phi}$ in 2D and as $\tau_{\phi}^{1/2}$ in
quasi-1D (many-channel wire) systems \cite{aa}, 
and thus diverges with lowering $T$, signaling the occurrence of the
strong localization (SL) regime. This prompts a question as to how the
system conducts at low $T$. 

For the case of electron-phonon scattering the answer is
well known. The conductivity is then governed by Mott's
variable-range hopping (VRH) \cite{mott}, yielding
$\sigma(T)\propto \exp\{-(T_0/T)^{\mu}\}$ with $\mu = 1/(d+1)$, where
$d$ is the spatial dimensionality. In the
presence of a long-range Coulomb interaction, the Coulomb gap in the
tunneling density of states modifies the VRH exponent, $\mu={1\over 2}$
\cite{efros-shklovskii}. 

But what is the low-$T$ behavior of $\sigma(T)$ if the
electron-phonon coupling is negligibly weak and the only source of
the inelastic scattering is the e-e interaction? Our purpose here
is to solve this long-standing fundamental problem, which is
also of direct experimental
relevance; see, e.g., Refs.~\cite{gershenson} and
\cite{hsu95,khondaker99}, 
where the crossover from WL to SL with lowering
$T$ was studied for
1D and 2D systems, respectively. For
definiteness, we concentrate on the case of a many-channel 1D
system with a short-range interaction. 
Our results are, however, more general 
(including single-channel wires, 2D systems, Coulomb interaction), as
we discuss in the end of the paper. 

It was proposed in \cite{fleishman78} that the e-e
interaction by itself is sufficient to induce VRH at
low $T$. This idea was widely used for interpretation of 
experimental \cite{khondaker99,shlimak99} and numerical
\cite{berkovits99}  results on 2D systems. 
Further, Ref.~\onlinecite{nattermann03} used 
bosonization to study the problem in 1D and concluded 
that transport is of VRH character.
These results are, however, in
conflict with the  argument \cite{fleishman80} -- 
supported by our analysis -- that elementary hops in the low-$T$ limit
are forbidden for $d<3$ even for the case of long-range ($1/r$) 
Coulomb interaction, since energy conservation cannot  
be respected when an electron attempts a real
transition by exciting an electron-hole pair \cite{malinin04}. The situation
is particularly interesting in 1D and 2D, where no mobility edge
exists, activation to which otherwise might give $\sigma (T)\neq 0$. 
If neither VRH nor activation, then what? 

Let us now specify the model. We consider a many-channel weakly
disordered wire, so that the relevant length scales satisfy $k_F^{-1}
\ll l \ll \xi$, where $k_F$ is the Fermi momentum, $l$ the mean free
path, and $\xi\sim \pi\nu D$ the localization length ($\nu$ is the
density of states per unit length and $D$ the diffusion
constant) \cite{efetov,adm-review}. 
The corresponding energy scales are the Fermi energy $E_F$,
the elastic scattering rate $\tau^{-1}$, 
and the level spacing in the localization
volume, $\Delta_\xi=1/\nu\xi$, with $E_F \gg \tau^{-1} \gg
\Delta_\xi$. We will assume a short-range interaction 
$U(\bf r-\bf r')$ between
electrons, characterized by a dimensionless coupling
$\alpha=\nu\tilde{U}(0)$, where $\tilde{U}({\bf q})$ is the Fourier
transform of $U({\bf r})$. 
We assume that $\alpha\ll 1$, which yields a richer behavior of
$\sigma(T)$ and allows better understanding of underlying physics; 
the case $\alpha\sim 1$ (as well as Coulomb interaction) 
will be discussed in the end.

At sufficiently high $T$, the conductivity
$\sigma(T)\simeq\sigma_{\rm D}+\Delta\sigma_{\rm WL}+\Delta\sigma_{\rm AA}$  is
close to its Drude 
value $\sigma_{\rm D}$, with quantum corrections related to the weak
localization ($\Delta\sigma_{\rm WL}$) and to interplay of interaction and
disorder (Altshuler-Aronov contribution $\Delta\sigma_{\rm AA}$)
\cite{aa}, 
\bea
&& {|\Delta\sigma_{\rm WL}|\over \sigma_{\rm D}} \sim
\int_{l_\phi^{-1}}{dq\over \pi\nu Dq^2} \sim {l_\phi\over \xi}
\sim \left({\Delta_\xi\over\alpha^2 T}\right)^{1/3}.
\label{e1}
\eea
Here we used the result for the dephasing rate length
$l_\phi=(D\tau_\phi)^{1/2}$ due to  e-e interaction \cite{aa},
\be
\label{e3}
\tau_\phi^{-1} \sim \alpha^2 T \int _{l_\phi^{-1}} {dq \over \pi\nu D
q^2} \sim \alpha^2 T {l_\phi\over \xi}.
\ee
The WL correction grows with lowering $T$ and finally becomes strong
($\Delta\sigma_{\rm WL}/\sigma_0\sim 1$) when $l_\phi$ reaches $\xi$, or,
equivalently, when $\tau_\phi^{-1} \sim \Delta_\xi$. This happens at
$T\sim T_1 = \alpha^{-2}\Delta_\xi$, marking the
beginning of the SL regime.  The
interaction-induced correction $\Delta\sigma_{\rm AA}/\sigma_{\rm D}\sim
(\alpha^2\Delta_\xi/T)^{1/2}$ remains small at
$T\sim T_1$ and thus is of no relevance in the present context. (For
$\alpha\sim 1$,  $\Delta\sigma_{\rm AA}$ is of order $\sigma_{\rm D}$
at $T\sim T_1$ and does not lead to any qualitative changes either.)
The subject of our interest is $\sigma(T)$ for $T<T_1$. 

\begin{figure}[ht]
\centerline{
\includegraphics[width=8cm]{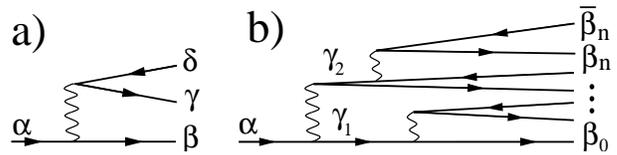}} 
\caption{Diagrams for the golden rule (a) and higher-order decay
  amplitude (b).}
\label{fig1} 
\end{figure}

In fact, SL does not necessarily mean $\sigma(T)$ is exponentially
small. Specifically, in the high-$T$ part of the SL regime the
transport mechanism -- we will call it power-law hopping (PLH) -- is
analogous to the one identified in \cite{gogolin75} for the case
of inelastic electron-phonon scattering, i.e., hopping over length 
$\sim\xi$ in time $\sim\tau_\phi$.  
In other words, the dephasing time $\tau_\phi$ serves in this regime 
as a lifetime of localized
states, which adds an imaginary part $i/2\tau_\phi$ to the single-particle
energies $\epsilon_\alpha$. This yields 
\be
\label{e4}
\sigma(T) \sim \sigma_{ac}(\Omega)|_{\Omega = i/\tau_\phi}\sim
e^2\nu \xi^2/\tau_\phi,
\ee
where $\sigma_{\rm ac}(\Omega)$ is the zero-$T$ conductivity of
noninteracting electrons at frequency $\Omega$.
The crucial point here is that $\tau_\phi$ in this SL regime can still be
calculated via Fermi's golden rule, as we are going to show. 
The lowest-order decay process of a localized state $|\alpha\rangle$  
is the transition to a
three-particle state -- two electrons $|\beta\rangle$,
$|\gamma\rangle$ and a hole $|\delta\rangle$, all located
within a distance $\sim \xi$, Fig.~\ref{fig1}a.
The corresponding matrix element of the interaction  is
\be
\label{e5}
V_{\alpha\beta\gamma\delta} \equiv V \sim \alpha\Delta_\xi; \qquad
|\epsilon_\alpha-\epsilon_\beta|, \  
|\epsilon_\gamma-\epsilon_\delta|\lesssim \Delta_\xi
\ee
and decays fast for larger energy differences (cf. results
for a metallic sample \cite{agkl,adm-review} with the dimensionless
conductance set to $g\sim 1$). Since 
energies of all the relevant single-particle states are within the
window of width $\sim T$, the level
spacing of three-particle states to which the original state
$|\alpha\rangle$  is coupled according to (\ref{e5}), reads
\be
\label{e6}
\Delta_\xi^{(3)} \sim \Delta_\xi^2/T.
\ee
Using (\ref{e5}), (\ref{e6}) and the golden rule, we find
\be
\label{e7}
\tau_\phi^{-1} \sim  |V|^2/\Delta_\xi^{(3)}
\sim \alpha^2 T.
\ee
Note that this result could also be obtained from (\ref{e3}) if
$\xi^{-1}$ is used as the infrared cutoff, as
appropriate for the SL regime. The condition of validity of the
golden-rule calculation is $V \gg \Delta_\xi^{(3)}$, or,
equivalently,  $\tau_\phi^{-1} \gg \Delta_\xi^{(3)}$. This introduces a new
temperature scale $T_3=\alpha^{-1}\Delta_\xi$, so that the PLH
regime is restricted to the range $T_3 \ll T \ll T_1$.
Combining (\ref{e4}) and (\ref{e7}), we get the conductivity in this
regime, 
\be
\label{e8}
\sigma(T) \sim e^2\nu\xi^2 \alpha^2 T \sim e^2\xi T/T_1.
\ee

\begin{figure}
\centerline{
\includegraphics[width=8.2cm]{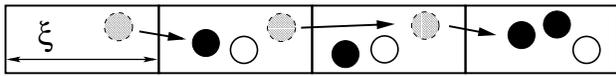}} 
\caption{Scheme of a high-order ballistic 
process: the string of particle-hole pairs (black and white
circles). The initial state and intermediate virtual states
($\gamma_i$) are shown in grey.}
\label{fig2} 
\vspace*{-0.2cm}
\end{figure}

What happens below
$T_3$? Simple hops on a distance $\sim\xi$ are then not sufficient to
delocalize electrons. Increasing the distance $r$ of a hop does  not help:  
the matrix element vanishes exponentially with  $r/\xi$, while the
level spacing $\Delta^{(3)}$ only as a power law. We thus have to analyze 
higher-order processes by exploring  the structure of
the theory in the many-body Fock space, similarly to the ideas developed 
for the problem of a quasiparticle decay in quantum dots
\cite{agkl,mf97,silvestrov}.  The process of $n$-th order represents a
transition with excitation of $n$ electron-hole pairs, $|\alpha\rangle
\longrightarrow |B^{(n)}\rangle\equiv |\beta_0\beta_1\bar{\beta}_1
\ldots\beta_n\bar{\beta}_n\rangle$ [with the energy difference $\lesssim
\Delta_\xi$ for each pair in view of (\ref{e5})], see  Fig.~\ref{fig2}b.
Let us estimate the dimensionless coupling of
$n$-th order (ratio of the matrix element to the level spacing of final
states), $V^{(n)}/\Delta^{(2n+1)}$, 
which is the $n>1$ generalization of the ratio
$V/\Delta_\xi^{(3)}$ considered above. In this estimate, it will be sufficient
for us to keep track of factors of the type $n^n$  (or $n!$) and $(T/T_3)^n$. 
Factors of the type $c^n$, where $c\sim 1$, will be
unessential and thus neglected. 

Let us assume that the set of states $B^{(n)}$
is spread over the length $m\xi$, with $\sim n/m$ pairs in each box of 
length $\xi$; later we optimize with respect to $m$ (for $1\lesssim m
\lesssim n$).  
The corresponding level spacing can be estimated as 
\be
\label{e9}
\Delta_{m\xi}^{(2n+1)}/\Delta_\xi 
\sim [(n/m) \Delta_{\xi}/T ]^n.
\ee
The matrix element 
$V_{\alpha;\beta_1,\ldots,\beta_{n+1}}\equiv V^{(n)}$ is given by
\be
\label{e10}
V^{(n)} = \sum_{\rm diagrams} 
\sum_{\gamma_1,\ldots\gamma_{n-1}} V_1 \prod _{i=1}^{n-1} 
{V_{i+1}\over  E_i-\epsilon_{\gamma_i}},
\ee
where the matrix element  $V_i$ corresponds
to the $i$-th interaction line, $\gamma_i$ are the virtual states
corresponding to internal lines, and $E_i$ the corresponding energy
variable which can be expressed as  
a linear superposition of $\epsilon_{\alpha_i}$ 
using the energy conservation. We need to take into account only those
contributions where all states forming each matrix element are within a
distance  $\sim\xi$ from each other, so that $V_i\sim\alpha\Delta_\xi$. 
Therefore, the summation over
each $\gamma_i$ is effectively taken over a single localization domain,
and we can replace it by taking the ``optimal'' $\gamma_i$, with  
$|E_i-\epsilon_{\gamma_i}|\sim\Delta_\xi$. We thus get 
\be
\label{e11}
V^{(n)}/\Delta_\xi \sim  [M_m^{(n)}]^{1/2} (\alpha\Delta_\xi)^n, 
\ee
where $M_m^{(n)}$ is the number of diagrams contributing to the
amplitude of the transition $|\alpha\rangle \to
|B^{(n)}\rangle$. These contributions have random
signs, hence the factor  $[M_m^{(n)}]^{1/2}$.  
  
To find  $M_m^{(n)}$, we first calculate  the number of topologically
different diagrams,  $D^{(n)}$, which satisfies the recursion relation
\be
\label{e12}
D^{(n)}=\sum_{n_1+n_2+n_3=n-1;\ n_{1,2,3}\ge 0} D^{(n_1)}D^{(n_2)}D^{(n_3)},
\ee
with the initial condition $D^{(0)}=1$.
It is easy to show that its solution increases only as $D^{(n)}\sim a^n$, with
$a\sim 1$, so that with the required accuracy it can be replaced by
unity. Thus $M_m^{(n)} \sim A_m^{(n)}$,  the number of allowed permutations of
the set $B^{(n)}$ over the final-state lines.
To estimate $A_m^{(n)}$, we notice that only electron-hole 
pairs within the same (or nearby) localization volume can be
interchanged, which yields $A_m^{(n)}\sim [(n/m)!]^{m} \sim (n/m)^{n}$.
Combining this with (\ref{e9}), (\ref{e11}), we finally get 
\be
\label{e14}
{V^{(n)} / \Delta^{(2n+1)}}\sim 
\left[\alpha(m/n)^{1/2} T / \Delta_\xi \right]^n.
\ee

Let us now analyze the result.
First of all, for given $n$ 
the most favorable case is $m\sim n$, which corresponds
to ``ballistic'' paths. In such a process, an electron makes a many-body
transition over the distance $n\xi$, leaving behind a string of $n$
particle-hole pairs, 
as illustrated in Fig.~\ref{fig2}. Second, 
$V^{(n)} / \Delta^{(2n+1)}$ increases with $n$ at sufficiently high $T$
and decreases with $n$ (thus remaining small for all $n$) at low
$T$. Therefore, at low $T$ the higher-order processes do not help a
localized single-particle state to decay, so that $\sigma(T)$ is exactly
zero. In contrast, at high $T$, the increase of the coupling
(\ref{e14}) with $n$ guarantees that the golden-rule calculation performed
for the WL and PLH regimes is not spoiled by the higher-order
effects. The temperature $T_c$ of the transition into the
zero-conductivity regime can be immediately estimated from (\ref{e14}),
$T_c\sim \Delta_\xi/\alpha$. In fact, (\ref{e14}) misses a
$\ln\alpha^{-1}$ factor recovered in a more accurate treatment below,
Eq.~(\ref{e16}).

What helps us to analyze the critical behavior at
the transition is that the structure of the theory, 
when restricted to the optimal (``ballistic'')
paths, reduces essentially 
to that of the Anderson model on the Bethe lattice -- a tree 
with a fixed branching number. Indeed, consider the  process shown in
Fig.~\ref{fig2}: 
an electron hops to an adjacent localization volume, creating an
electron-hole pair, then to the next one and so forth. Clearly, the density of
final states increases at each step by the same factor 
\be 
\label{e15}
K\sim \Delta_\xi/ \Delta_\xi^{(3)} \sim T/\Delta_\xi,
\ee
which is the branching number of the Bethe lattice. 
The Bethe-lattice character of the problem can also be
inferred from the exponential dependence of the coupling
(\ref{e14}) on $n$ at $m=n$. This should be contrasted with the opposite
limit, $m=1$, corresponding to the case of a quantum dot, when (\ref{e14})
contains an additional $n^{-n/2}$ factor. The latter is related to a
decrease of the 
effective branching number with $n$ in this case, as was noticed in
\cite{mf97,silvestrov}. In other words, the mapping on the Bethe
lattice model, oversimplified for a quantum dot
(for which it was originally proposed in \cite{agkl}), turns out to work
perfectly in the case of localized states in a wire (or, more generally, in a 
non-restricted geometry), where going to higher generations in Fock space
can be accompanied by the exploration of new regions in real space.  

We can now use the results for the Anderson transition on the Bethe lattice
that has been studied extensively
\cite{abou73,efetov,zirnbauer86,mf94,mf97}. For a large branching
number $K$ the equation for the transition point reads 
\be
\label{e16}
\Delta/V = 4 \ln K,
\ee
where $V$ is the hopping matrix element, and $\Delta$ is the mean 
level spacing of
states of generation $n+1$ coupled to a given state of generation $n$. 
Using (\ref{e5}), (\ref{e15}), and $\Delta=\Delta_\xi^{(3)}$, we find
the transition temperature,
\be
\label{e17}
T_c \sim \Delta_\xi/\alpha \ln \alpha^{-1},
\ee
so that $T_c\sim T_3/\ln \alpha^{-1}$.
The critical behavior of $\sigma(T)$ above $T_c$ is governed
by that of the decay rate $\tau_\phi^{-1}$, which translates into the imaginary
part of the self-energy ${\rm Im}\:\Sigma$ 
for the Bethe-lattice problem. The critical behavior
of the latter was found  in \cite{mf94,mf97}, yielding
\be
\label{e18}
\sigma(T) \propto {\rm Im}\: \Sigma \propto
\exp \{- c_\alpha [(T-T_c)/T_c]^{-1/2} \},
\ee
with $c_\alpha \sim \ln\alpha^{-1}$ for $\alpha\ll 1$. 
Near the transition, the local density of states on the Bethe lattice acquires
an increasingly more sparse ``spatial'' structure \cite{mf94,mf97}, 
so that the transport is
governed by  processes connecting remote states in Fock space, 
Fig.~\ref{fig2}, which implies a glassy character of the system. 
When $T\to T_c$, the length of the particle-hole strings diverges.
We term the low-$T$ phase ``Anderson-Fock glass" (AFG), since 
its physics is governed by the Anderson localization in Fock space.

\begin{figure}[ht]
\centerline{
\includegraphics[width=8cm]{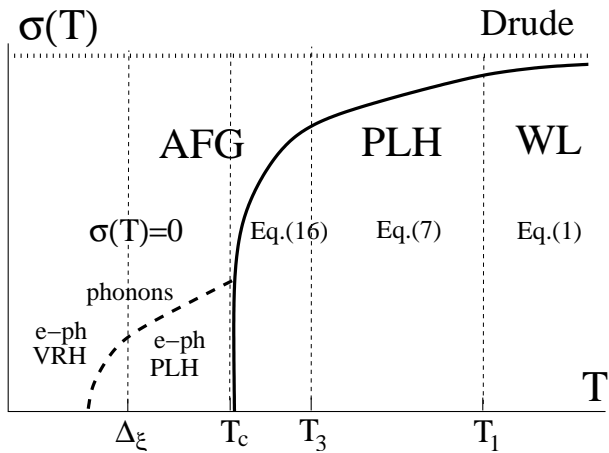}} 
\caption{Schematic behavior of $\sigma(T)$ on the log-log scale: 
the weak-localization (WL), power-law
hopping (PLH), and Anderson-Fock glass (AFG) regimes, with a localization
transition at $T_c$. Dashed line: PLH and VRH contribution to $\sigma(T)$ in
the case of a weak coupling to phonons. }
\vspace*{-0.3cm}
\label{fig3} 
\end{figure}

The found behavior of $\sigma(T)$ is illustrated in Fig.~\ref{fig3}.   
If $\alpha\sim 1$, all scales become of the same order, $T_1\sim T_3\sim
T_c\sim \Delta_\xi$, and the range of PLH disappears. 
In realistic systems, weak coupling to phonons will lead to PLH 
(for $\alpha\ll 1$) and VRH below $T_c$, 
but with a small prefactor, so that the
transition at $T_c$ should be well observable.  
Note the peculiar character of the transition: not only the exponential
critical behavior (\ref{e18}) but also that the ordered (metallic)
phase corresponds to $T>T_c$. An apparent conflict with the
Mermin-Wagner theorem is related to the unconventional (functional)
nature of the order parameter for Anderson localization
\cite{efetov,zirnbauer86,mf94}.  

Before closing the paper, let us briefly mention a few extensions of our
results \cite{tobe}. 

(i) {\it 1D: Single channel.} 
We have recently shown \cite{luttinger-dephasing} that the notion of
WL  and dephasing are also applicable to a disordered Luttinger liquid 
and calculated $\sigma(T)$ in the WL regime. 
The low-$T$ results presented above can then be easily
generalized to the single-channel case. An important difference
is that $\tau$ (and thus, $\xi$)
is strongly renormalized by interaction, $\tau(T)/\tau\sim
(T/E_F)^{\alpha}$. 

(ii) $d>1$.
Generalization to 2D systems is straightforward, with $\xi$ depending 
exponentially on disorder,
$\xi\propto \exp(2\pi^2\nu D)$. Our consideration is also applicable
to 3D Anderson insulators; however, in this case $\sigma(T<T_c)$ 
will not be exactly zero but rather determined by
the activation above the single-particle mobility edge.

(iii) {\it Coulomb interaction.} 
For 1D and 2D geometry, the transition survives also for $1/r$
Coulomb interaction, since correlated hops of two electrons
separated by a large distance $r\gg\xi$ will not delocalize them. 
Indeed, the
corresponding matrix elements decrease with $r$ as
$V_{\alpha\beta\gamma\delta}\propto 1/r^3$, which is not compensated
by the increase $\propto r^d$ of the density of final states for $d<3$
and thus does not help. On the other hand, in 3D such processes will lead to
delocalization for any $T$ \cite{fleishman80}; the behavior of
$\sigma(T)$ in this case requires a separate study.

(iv) {\it Creep.} Our approach can be used to
analyze the non-linear conductivity $\sigma(E)$ 
at weak electric field $E$. 

In conclusion, we have studied the conductivity of interacting
electrons in a disordered quantum wire; very similar results hold for
a 2D system. In contrast to a popular belief, the e-e interaction is
not sufficient to support the VRH transport in the low-$T$
limit. Instead, the system undergoes a localization transition at
temperature  $T_c$, below which $\sigma(T)=0$ (assuming vanishing
coupling to phonons).  We have shown that the conductivity vanishes as
$\ln\sigma(T)\propto - (T-T_c)^{-1/2}$ as $T\to T_c$. Transport in the
critical regime is governed by  many-particle transitions between
distant states in Fock space, corresponding to the formation of long
strings of electron-hole pairs. For weak interactions, this
Anderson-Fock glass phase is separated from the WL regime by an
intermediate temperature regime of power-law hopping.

We thank V.~Cheianov, D.~Maslov, G.~Minkov, T.~Nattermann, and B.~Shklovskii
for valuable discussions. We are particularly grateful to 
I.~Aleiner, B.~Altshuler, and D.~Basko for very useful discussions and
criticism of the earlier version of this work (cond-mat/0407305v1).
This helped us to correct an error in 
counting of the number of diagrams $M_m^{(n)}$ [which yielded a spurious
double exponential tail of $\sigma(T)$ in
the AFG phase], whose elimination gave a true phase transition at
$T=T_c$, as also found in \cite{aleiner}.
The work was supported by SPP ``Quanten-Hall-Systeme'' and CFN 
of the DFG and by RFBR.

\end{document}